\documentclass[12pt]{article}
\usepackage{latexsym}
\newcommand{\be}{\begin{equation}}
\newcommand{\ee}{\end{equation}}
\def\n{\noindent}
\catcode `\@=11 \catcode `\@=12
\begin{document}
\begin{center}

{\bf {Analytical Solution For Navier-Stokes Equations In Two Dimensions For Laminar Incompressible Flow}} \\
\vspace{5mm} \normalsize{Saeed Otarod
$^{\ast}$\footnote{Corresponding Author}and Davar Otarod $^{\dag}$
} \\
\vspace{10mm} \normalsize{$^{\ast}$\it{Department of Physics, Yasouj
University, Yasouj, Iran \\
E-mail : sotarod@mail.yu.ac.ir, sotarod@yahoo.com}}\\
\vspace{5mm} \normalsize{$^{\dag}$\it{ Department of mechanical
engineering, Yasouj University,
Yasouj, Iran}} \\

\end{center}
\vspace{10mm}

\begin{abstract}
The Navier-Stokes equations describing laminar flow of an
incompressible fluid will be solved. Different group of general
solutions for Navier stokes equations governing Laminar
incompressible fluids will be derived.
\end{abstract}
\smallskip
\n Keywords: {Applied mathematics ,  Fluid dynamics, Navier-Stokes
equations}
\section{Introduction}
Navier stokes equations for laminar flows, in different physical
conditions, have been solved in almost all fluid mechanics text
books(1). In some physical conditions these equations have exact
solutions, but in general these equations  are complicated
nonlinear partial differential equations that can not be solved
easily. Therefore, the authors in most cases, in order to solve
these equations, have either used numerical methods, or they have
appealed to Physically acceptable simplifying considerations. In
this way so many efforts have been put on the recognition of
physical parameters and terms that could be neglected without
affecting the system effectively.\\
Also, Navier stokes equations have a very
wide application in stellar astrophysics,[2],[3],[4].\\
 Since 2000, The author has tried to somehow overcome the complexities
governing the analytical solutions of nonlinear partial
Differential equations and in many respects
these activities has been successful.[5],[6],[7],[8],[9],[10].\\
 Since Navier stokes equations governing the
incompressible fluids has a wide application in technology and
industry, in this article we try to find explicit solutions of
these equations  for the laminar incompressible fluids in two
dimensions. We have treated these set of equations in a general
way without restricting ourselves to any especial boundary value
conditions, although from the results found in sections 2 and 3,
one can conclude that these results are consistent with different,
logical and acceptable, boundary conditions. How to fit the
results with a especial set of boundary conditions is beyond the
scope of this article. We will address this issue in our next
efforts.
\section{Basic equations}
 Navier stokes equations in two dimensions in cartesian
 coordinates for laminar incompressible fluids are;
\begin{equation}
u(x,y)\frac{\partial u(x,y)}{\partial x}+v(x,y)\frac{\partial
u(x,y)}{\partial y}=-\frac{1}{\rho}\frac{\partial}{\partial
x}P(x,y)+\nu(\frac{\partial^2 u(x,y)}{\partial
x^2}+\frac{\partial^2 u(x,y)}{\partial y^2})
\end{equation}
\begin{equation}
u(x,y)\frac{\partial v(x,y)}{\partial x}+v(x,y)\frac{\partial
v(x,y)}{\partial y}=-\frac{1}{\rho}\frac{\partial}{\partial
y}P(x,y)+\nu(\frac{\partial^2 v(x,y)}{\partial
x^2}+\frac{\partial^2 v(x,y)}{\partial y^2})
\end{equation}
\begin{equation}
\frac{\partial u(x,y)}{\partial x}+\frac{\partial v(x,y)}{\partial
y}=0
\end{equation}
Here, $u(x,y)$ and $v(x,y)$ are the $x$ \hskip .1cm and $y$
components of the velocity respectively, and $P(x,y)$ stands for the
pressure. The two first equations are the components of momentum
equations, in the $x $\hskip .1cm and $y$ direction. and obviously
the third equation is the
continuity equation. \\

 To solve the above set of equations, at first we suppose that, $u $
 and $v$, are functions of a float function $f(x,y)$, that is to
 say,
\begin{equation}
u=u(f(x,y)),v=v(f(x,y))
\end{equation}
However, this is a restricting assumption, it has no contradiction
with logical considerations based on the following discussions.\\
1-What we do is similar to the application of separation method in
solving the problems. At the beginning, when we apply this method
we do not have any logical supportive argument behind that. We
hope by writing the function in the separated form, we will come
to suitable results. Of course, It is possible that it may not
work and in fact in many cases it does not. If it works  we
have succeeded, otherwise it would be put aside.\\
 2-Here,What we do is to collect all the variables into a single
 float variable $f$. In fact, it is Some sort of variable change or
 transformation.\\
3- Any way $u$ and $v$, some how are related to each other, in
advance we do not know this relationship, one possibility is that,
they both are the function of  the same function. this is the case,
in many physical situations, for example in electromagnetic
radiation all components of electrical and magnetic fields all are
functions of $K.r +\omega t$ and in harmonic oscillation the
oscillations in all directions are all functions of $\omega t$
.Therefore it is reasonable to think that the two components of the
velocity are the functions of a common function like $f(x,t)$.\\
Now according to the above considerations equation(3) will be
written as;
\begin{equation}
\frac{d u}{d f}\frac{\partial f}{\partial x}+\frac{d v}{d
f}\frac{\partial f}{\partial y}=0
\end{equation}
Since $f$ is a float and arbitrary function, so far as it fulfills
the conditions expressed by the equations 1- 4,any properties can
be attributed to that. Obviously at first we prefer to chose $f$
in such a way that the equations  could be solved as easily as
possible. We do have so Many options and as the first choice we
assume
\begin{equation}
 \beta
\frac{\partial f}{\partial x}=\alpha \frac{\partial f}{\partial y}
\end{equation}
with this assumption equation(5) will be simplified to
\begin{equation}
\alpha \frac{du}{df}+\beta \frac{dv}{df}=0
\end{equation}
which immediately results in
\begin{equation}
\alpha u(f)+\beta v(f)=c
\end{equation}
As we mentioned this is not the only possible choice and for example
we may choose $f$ to be
\begin{equation}
\frac{\partial f}{\partial x}=f \frac{\partial f}{\partial
y}\end{equation} or\\
\begin{equation} \frac{\partial f}{\partial x}=e^{f}\frac{\partial
f}{\partial y}
\end{equation}
What choice is the  most suitable one, it depends on the boundary
conditions that are to be satisfied, at this point let us
concentrate on equations (6) and (7) and look for the consequences.
From elementary algebra The solution to equation(6) is ,
\begin{equation}
f=f(\alpha x+\beta y)
\end{equation}
That is to say $f$ can be any arbitrary function of $(\alpha x+\beta
y)$. From equations(7) and (8) we have
\begin{equation}
\frac{\partial v}{\partial x}=-\frac{\alpha}{\beta}\frac{\partial
u}{\partial x}, \hskip 2cm \frac{\partial^2 v}{\partial
x^2}=-\frac{\alpha}{\beta}\frac{\partial^2 u}{\partial x^2}
\end{equation}
Substituting results of (11) into equation(2), we will come to
\begin{equation}
u(x,y)\frac{\partial u(x,y)}{\partial x}+v(x,y)\frac{\partial
u(x,y)}{\partial
y}=\frac{\beta}{\alpha}\frac{1}{\rho}\frac{\partial}{\partial
y}P(x,y)+\nu(\frac{\partial^2 u(x,y)}{\partial
x^2}+\frac{\partial^2 u(x,y)}{\partial y^2})
\end{equation}
subtracting equation(1) from equation(12) will result in
\begin{equation}
    \frac{1}{\rho}(\alpha \frac{\partial P}{\partial x}+\beta \frac{\partial P}{\partial
    y})=0
\end{equation}
The general solution to this equation is $P=P(\beta x-\alpha y)$,
that is, P can be any function of $(\beta x-\alpha y)$.\\Now if we
rewrite equation(1) in this way;
\begin{equation}
u(x,y)\frac{\partial u(x,y)}{\partial x}+v(x,y)\frac{\partial
u(x,y)}{\partial y}-\nu(\frac{\partial^2 u(x,y)}{\partial
x^2}+\frac{\partial^2 u(x,y)}{\partial
y^2})=-\frac{1}{\rho}\frac{\partial}{\partial x}P(x,y)
\end{equation}
it will be seen that, while according to our first assumption the
left hand side of the above equation is a function of $f(\alpha
x+\beta y)$ the right hand side has to be a function of $(\beta
x-\alpha y)$, but these functions are linearly independent. This can
not happen unless $\frac{\partial P}{\partial x}=0$ or
$\frac{\partial p}{\partial x}=constant$. The same argument is true
for y dependence of $P$; that is, $\frac{\partial P}{\partial y}=0$
or $\frac{\partial P}{\partial y}=constant$. In this way we find
that our original assumption$ u=u(f)$ and $v=v(f)$ is only
consistent with those physical situations, at which
\begin{equation}
p=constant
\end{equation}
or
\begin{equation}
p=\beta x - \alpha y
\end{equation}
So far we have derived two answer for $P$, and each will lead to a
different result. If we follow our analysis by taking into account
the result of equation (15) and substituting that into the equations
(1) and (2), we will see that both of these equations are identical.
Therefore we need only one of them to be solved. As a result
equation(1) will be written as
\begin{equation}
u(x,y)\frac{\partial u(x,y)}{\partial x}+v(x,y)\frac{\partial
u(x,y)}{\partial y}=\nu(\frac{\partial^2 u(x,y)}{\partial
x^2}+\frac{\partial^2 u(x,y)}{\partial y^2})
\end{equation}
since $u=u(f)$,the following results are valid
\begin{equation}
\frac{\partial u}{\partial x}=\frac{\partial f}{\partial x}
\frac{du}{df} \hskip 1cm\frac{\partial u}{\partial
y}=\frac{\partial f}{\partial y} \frac{du}{df}
\end{equation}
\begin{equation}
\frac{\partial ^2u}{\partial x^2}=\frac{d^2 u}{d
f^2}(\frac{\partial f}{\partial x})^2+\frac{\partial ^2
f}{\partial x^2}\frac{d u}{d f} \hskip 1cm \frac{\partial
^2u}{\partial y^2}=\frac{d^2 u}{d f^2}(\frac{\partial f}{\partial
y})^2+\frac{\partial ^2 f}{\partial y^2}\frac{d u}{d f}
\end{equation}
From equations(6), (8), (19) and (20)  will will have
\begin{equation}
\frac{c}{\beta} \frac{d u}{d f} \frac{\partial f}{\partial
y}=\nu(\frac{d^2 u}{d f^2}[(\frac{\partial f}{\partial
x})^2+(\frac{\partial f}{\partial y})^2]+\frac{d u}{d
f}[\frac{\partial^2 f}{\partial x^2}+\frac{\partial^2 f}{\partial
y^2}])
\end{equation}
Up to this point $u $  has been separated from other parameters. To
be able to solve for $u$ we have to decide about $ f $. In the next
two subsections we will try two different cases.
\subsection{$f=\alpha x+ \beta y $} As we said $f$ can be any function of $\alpha x+ \beta y $
and each $f$ will give results that are consistent with an
especial boundary conditions. The simplest choice will be
\begin{equation}
f=\alpha x+ \beta y
\end{equation}
For this choice, equation(20) will be reduced to
\begin{equation}
\frac{c}{\nu(\alpha^2+\beta^2
)}\frac{d^2u}{df^2}=\frac{d^2u}{df^2}
\end{equation}
The obvious solution to this differential equation is
\begin{equation}
u=Ae^{\frac{c}{\nu(\alpha^2+\beta^2)}(\alpha x+\beta y)}+B
\end{equation}
where A and B are constants of integration. Substituting the above
result into the equation(8) $v$ will be found out to be;
\begin{equation}
v(f)=\frac{1}{\beta}(c-\alpha B-\alpha
Ae^{\frac{c}{\nu(\alpha^2+\beta^2)}(\alpha x+\beta y)})
\end{equation}
\subsection{$f=e^{\alpha x+\beta y}$}
For this $f$ we will have;\\
 \begin{equation}\frac{\partial f}{\partial x}=\alpha
f,\hskip .2 cm\frac{\partial^2 f}{\partial x^2}=\alpha^2 f,\hskip 1
cm\frac{\partial f}{\partial y}=\beta f,\hskip .2 cm
\frac{\partial^2 f}{\partial y^2}=\beta^2 f\end{equation}
consequently equation(20) will be
\begin{equation}
(\frac{c}{\nu(\alpha^2+\beta^2)}-1)\frac{du}{df}=f\frac{d^2u}{df^2}
\end{equation}
which has the solution
\begin{equation}
u=A+B f^{\frac{c}{\nu(\alpha^2+\beta^2)}}
\end{equation}
or in terms of $x $\hskip .1cm and $y $ it will be
\begin{equation}
u=A+B (\alpha x+\beta y)^{\frac{c}{\nu(\alpha^2+\beta^2)}}
\end{equation}
Accordingly from equation(8)  $v$ will be
\begin{equation}
v=\frac{c-\alpha A}{\beta}-\alpha (\alpha x +\beta
y)^{\frac{c}{\nu(\alpha^2+\beta^2)}}
\end{equation}
Of course for each  choice of $f$ we will have a different
solution. But the solutions found in this way all have the simple
dependance on $\alpha x+\beta y$. Now we are looking for solution
that will have more complicated dependance on $ x$ and $ y$.This
helps us to find solutions that are consistent with other type of
boundary conditions. In the next section we will study this aspect
of the problem.
\section{Other solutions}

In this section the basic idea is that, having two different
particular solution, how we may find other solutions. Suppose we
select the two following particular solutions .
\begin{equation}
u_1=Ae^{\frac{c}{\nu(\alpha^2+\beta^2)}(\alpha x+\beta y)}+B
\end{equation}
Of course if we substitute $-\beta$ in place of $\beta$ in the
above equation the result will also be another linear independent
solution to our equations. Therefore our next choice will be.
\begin{equation}
u_2=Ae^{\frac{c}{\nu(\alpha^2+\beta^2)}(\alpha x-\beta y)}+B
\end{equation}
We have to remind that we are quite free to chose any particular
solution we like.\\ Now we look for those $u$ and $v$ and $P $ that
are functions of $u1$ and $u2$. That is to say :
\begin{equation}
u=u(u_1,u_2)\hskip .5cm, v=v(u_1,u_2),\hskip .5cm P=P(u_1,u_2),
\end{equation}
Substituting equations(31) in the continuity equation, (equation(3))
will lead to:
\begin{equation}
\frac{\partial u}{\partial u_1}\frac{\partial u_1}{\partial
x}+\frac{\partial u}{\partial u_2}\frac{\partial u_2}{\partial
x}+\frac{\partial v}{\partial u_1}\frac{\partial u_1}{\partial
y}+\frac{\partial v}{\partial u_2}\frac{\partial u_2}{\partial y}=0
\end{equation}
If we differentiate $u_1$ and $u_2$ with respect to $x$ and $y$ and
if we substitute the result in equation(32) we will have
\begin{equation}
e^{\frac{c}{\nu(\alpha^2+\beta^2)}(\alpha x+\beta y)}(\alpha
\frac{\partial u}{\partial u_1}+\beta \frac{\partial v}{\partial
u_1})+e^{\frac{c}{\nu(\alpha^2+\beta^2)}(\alpha x-\beta y)}(\alpha
\frac{\partial u}{\partial u_2}-\beta \frac{\partial v}{\partial
u_2})=0
\end{equation}
Since the exponentials are linearly independent, it is obvious that
we may write
\begin{equation}\
\alpha\frac{\partial u}{\partial u_1}+\beta \frac{\partial
v}{\partial u_1}=0
\end{equation}
\begin{equation}\
\alpha\frac{\partial u}{\partial u_2}+\beta \frac{\partial
v}{\partial u_2}=0
\end{equation}
Upon integration we will come to
\begin{equation}
\alpha u+\beta  v =G(u_2)
\end{equation}
\begin{equation}
\alpha u-\beta  v =H(u_1)
\end{equation}
Where $G$ and $H$ are arbitrary functions. From the above equations
we will come to
\begin{equation}
u=\frac{1}{2\alpha}(G(u_2)+H(u_1))
\end{equation}
\begin{equation}
v=\frac{1}{2\beta}(G(u_2)-H(u_1))
\end{equation}
As before, depending on what functions we choose for G and H, we
will come to different results, Suppose we choose $G=u_2 $ and $
H=u_1 $ therefore
\begin{equation}
u=\frac{1}{2\alpha}(u_1+u_2)
\end{equation}
\begin{equation}
v=\frac{1}{2\beta}(u_1-u_2)
\end{equation}

Substituting the above results in equations (1) and (2) we will come
to the following results for $P(x,y)$

\begin{equation}
{\frac {{\frac {\partial }{\partial x}}P \left( x,y \right)
}{\rho}}=1 /2\, \left( -2\,A{e^{{\frac {c \left( \alpha\,x+\beta\,y
\right) }{\nu \, \left( {\alpha}^{2}+{\beta}^{2} \right) }}}}+
\left( 1+{e^{2\,{ \frac {c\beta\,y}{\nu\, \left(
{\alpha}^{2}+{\beta}^{2} \right) }}}}
 \right)  \left( c-B \right)  \right) A{e^{{\frac {c \left( \alpha\,x-
\beta\,y \right) }{\nu\, \left( {\alpha}^{2}+{\beta}^{2} \right)
}}}}c {\alpha}^{-1}{\nu}^{-1} \left( {\alpha}^{2}+{\beta}^{2}
\right) ^{-1}
\end{equation}
\begin{equation}
{\frac {{\frac {\partial }{\partial y}}p \left( x,y \right)
}{\rho}}=1 /2\, \left( -c+B \right) Ac{e^{{\frac {c \left(
\alpha\,x-\beta\,y
 \right) }{\nu\, \left( {\alpha}^{2}+{\beta}^{2} \right) }}}} \left( -
1+{e^{2\,{\frac {c\beta\,y}{\nu\, \left( {\alpha}^{2}+{\beta}^{2}
 \right) }}}} \right) {\beta}^{-1}{\nu}^{-1} \left( {\alpha}^{2}+{
\beta}^{2} \right) ^{-1}
\end{equation}
If we take $B=c$ we will have
\begin{equation}
\frac {\partial }{\partial y}p =0
\end{equation}
and
\begin{equation}
\frac{\partial}{\partial}p=-\rho{A}^{2}{e^{2\,{\frac
{c\alpha\,x}{\nu\, \left( {\alpha}^{2}+{\beta}^{ 2} \right)
}}}}c{\alpha}^{-1}{\nu}^{-1} \left( {\alpha}^{2}+{\beta}^{2 }
\right) ^{-1}
\end{equation}
or
\begin{equation}
p=-1/2\,{A}^{2}{e^{2\,{\frac {c\alpha\,x}{\nu\, \left(
{\alpha}^{2}+{ \beta}^{2} \right) }}}}{\alpha}^{-2}+D
\end{equation}
Where D is a constant of integration.

As a result we may say that the set of functions $u$, $v$ and $P$
defined by equations(29),(30),(40),(41),(46) are  new generations of
the solutions to Navier stokes equations.

\section{Discussion}
The retrieved results show that there are an infinite number of
solutions to the Navier stokes equations. Any choice for $f(\alpha
x+\beta )y$ in equation(20) will result in a new solution. How to
chose $f$, depends on the Boundary
conditions to be applied.\\
Since all $f$'s in section 2 were all specifically functions of
$\alpha x+\beta y$ the resulting solutions for $u$ and $v$, will
also be only dependent on $\alpha x+\beta y$. Therefore they will,
be consistent to some especial set of boundary conditions. In
section 3, In order to increase the capabilities of the solutions,
we tried to find out solutions with different type of dependency on
$x$ and $y$. Mathematically ,what we did in that section is somehow
similar to what we do in linear scheme. In linear scheme we prove
that if $u_1$ and $u_2 $ are particular solutions of a linear
differential equation, then $u=c_1 u_1 +c_2 u_2$ \hskip .1cm will be
the general solution for that differential equation. In a similar
manner, in section 3, we showed that if $u_1$ and $u_2$ are
particular solutions  to Navier stokes equations then,
$u=\frac{1}{2\alpha}(G(u_2)+H(u_1))$ will be a general solution to
that equation. Since from the results of section 2 there are
numerous Particular solutions, available, therefore numerous set of
general solutions are predictable for our set of nonlinear partial
differential equations.\\
The results found this way in nonlinear regime is rather different
from what we come to, in linear discussion. In linear discussion, We
argue that if, $u_1$ and $u_2 $ are particular solutions, then the
only general solution will be $u=c_1 u_1 +c_2 u_2$, while here in
the recent discussion we come to the result that for any pair of
particular solutions we will have a new set of general solutions.
consequently one can not recognise from the beginning which set of
General solutions are consistent with the required boundary
condition.\\
To argue this point, is beyond the scope of this article. Here, What
we may suggest is that; It is possible to focus from the beginning,
on those float functions $f$ that are consistent  with all or part
of the boundary conditions and continue our calculations based
on those choices.\\
In short,The scheme we followed in this article provides us numerous
acceptable solutions, and it shows that with preliminary knowledge
of calculus analysis we can overcome the complexities of solving
nonlinear set of partial differential equations.\\
At the end, what we suggest as the topics of studies in future is;
"how we may consider the effects of  boundary conditions from the
beginning".

\section{acknowledgment}
The author thanks scranton University for providing him the
facilities during his Sabbatical leave, that helped him to conduct
this research. The author also would like to thank , Dr. Robert A.
Spalletta. and  Dr. Paul. Fahey, for their educating comments on the
subject.
\section{references}
1-R. Byron Birth., W. E. Stewart., E. N. Lightfoot., Transport
Phenomena. 1960. Jhon wiley and sons.\\
2- Field. G. B. 1965, {\em Ap. J.}, 142, 531.\\
3- Meerson, B., 1989, {\em Ap. J.}, 374, p. 1012.\\
4- Meerson, B. Megged, E., 1996, {\em Ap. J.}, 457, p. 321.\\
5- Otarod, S. Analytical solution for the Navier-Stokes equations,{
$26^{th}$ Intl. Coll. on Group
Theoretic Methods in Physics}, 26-30 July, 2006, New York.\\
6- Otarod, S. A New and Powerful Method for Solving Nonlinear
Partial Differential Equations. {\em Proceed. $24^{th}$ Intl. Coll.
on Group
Theoretic Methods in Physics}, 15-20 July, 2002, Paris\\
7- Otarod, S. An Explicit Solution to Hopf's Equation. {\em
Electronic J. Diff. Eq}, Prob. Section 2003-3. \\
8- Otarod, S.; Ghanbari, J.  Separation of Variables for Nonlinear
Diff. Eqs. {\em Electronic J. Diff. Eq}, Prob. Section 2002-2.\\
9- Otarod, S. Thermal instability of the interstellar matter for
different Heating/Cooling functions: Analytical solutions using the
method of separation of variables . PhD Thesis, Ferdowsi University,
Iran, 2001.\\
10- Ghanbari, J. and Otarod, S., Analytical solution of nonlinear
dynamical equations in Interstellar media at quasi hydrostatic
equilibrium. {\em Proc. M31-M33}, May 21-25, Germany, 2000.\\
\end{document}